\documentclass{article}
\usepackage{spconf,amsmath,graphicx}

\title{NOISE ROBUST IOA/CAS SPEECH SEPARATION AND RECOGNITION SYSTEM FOR THE THIRD 'CHIME' CHALLENGE}
\name{Xiaofei Wang, Chao Wu, Pengyuan Zhang, Ziteng Wang, Yong Liu, Xu Li, Qiang Fu, Yonghong Yan\thanks{This work is partially supported by the National Natural Science Foundation of China (Nos. 11161140319, 91120001, 61271426), the Strategic Priority Research Program of the Chinese Academy of Sciences (Grant Nos. XDA06030100, XDA06030500), the National 863 Program (No. 2012AA012503) and the CAS Priority Deployment Project (No. KGZD-EW-103-2).}}
\address{Institute of Acoustics, Chinese Academy of Sciences}
\begin{document}
%
\maketitle
\begin{abstract}
This paper presents the contribution to the third 'CHiME' speech separation and recognition challenge including both front-end signal processing and back-end speech recognition. In the front-end, Multi-channel Wiener filter (MWF) is designed to achieve background noise reduction. Different from traditional MWF, optimized parameter for the tradeoff between noise reduction and target signal distortion is built according to the desired noise reduction level. In the back-end, several techniques are taken advantage to improve the noisy Automatic Speech Recognition (ASR) performance including Deep Neural Network (DNN), Convolutional Neural Network (CNN) and Long short-term memory (LSTM) using medium vocabulary, Lattice rescoring with a big vocabulary language model finite state transducer, and ROVER scheme. Experimental results show the proposed system combining front-end and back-end is effective to improve the ASR performance.
\end{abstract}
\section{Introduction}
\label{sec:intro}
Automatic Speech Recognition (ASR) has been applied to many human-computer interaction systems, such as tablet computer, smartphones, personal computers and televisions. Meanwhile, robust ASR in noisy environments is paid more attention due to its applicable value. The 3rd 'CHiME' speech separation and recognition challenge is such a platform for testing the recognition rate of noisy speech in complex environments \cite{barker2015thethird}. Our contributions to CHiME are separated into two parts: front-end techniques and back-end techniques.

It is well known that a lot of front-end techniques aim at extracting clean desired speech signals. Among them, multi-channel system is proved effective to improve the front-end performance in noisy and reverberant environment so that it attracts more attention in consideration of better balance between noise reduction and speech distortion. As is known to all, more noise reduction doesn't mean more clean desired speech. Speech distortion brought by artifacts affects ASR performance severely. Therefore, taking speech distortion into account in the multi-channel optimization criterion, multi-channel wiener filter (WMF) technique has been proposed to estimate the desired speech component in noisy environment \cite{doclo2001gsvd}. The technique is generalized as speech distortion weighted MWF (SDW-MWF). The tradeoff between noise reduction and speech distortion is taken into consideration. In principle, it is desired to have less noise reduction in speech dominant segments and more noise reduction otherwise. From this motivation, we improve the SDW-MWF by focusing on the tradeoff parameter optimization from the perspective of desired noise reduction control technique.

Recently, acoustic modelling based on the Deep Neural Networks (DNNs) has gained popularity with the consistent improvement in recognition performance over earlier Neural Network based front-ends (e.g.\cite{Grezl2007}). DNNs are either deployed as the front-end for standard Hidden Markov Model based on Gaussian Mixture Models (HMM-GMMs), or in a hybrid form to directly estimate state level posteriors. As noted in several publications \cite{Larochelle2009, Seide2011, Swietojanski2013, Liu2014}, DNNs show general word error rate (WER) improvements on the order of 10-30\% relative across a variety of small and large vocabulary tasks when compared with HMM-GMMs built on classic features. A DNN is a conventional Multi-Layer Perceptron (MLP) with many internal or hidden layers. Convolutional Neural Networks (CNNs) are an alternative type of neural network that can be used to reduce spectral variations and model spectral correlations which exist in signals. CNNs are a more effective model for speech compared to DNNs \cite{Sainath2013Deep}. Besides, Long Short-Term Memory (LSTM) is also a specific recurrent neural network (RNN) architecture that was designed to model temporal sequences and their long-range dependencies more accurately than conventional RNNs. LSTM are also proved more effective than DNNs and conventional RNNs for acoustic modeling \cite{Sak2014Long, Sak2014Long2}. In this paper, we take advantage of these techniques for acoustic modeling and make a combination of them to achieve a better ASR performance \cite{Fiscus1997}.

This paper is organized as follows. In section 2, 3, we describe the front-end and back-end of the proposed system. In section 4, we carry out ASR experiments and list the results with analysis. At last, we draw a conclusion in section 5.

\section{Speech enhancement front-end}

\label{ssec:Signal Model}
In order to suppress background noise, multichannel wiener filter (MWF) is introduced to the multi-microphone set-up \cite{doclo2001gsvd}. Since MWF does not require transfer functions between a target speaker and microphones, it is suitable for the CHiME3 task.
Taking speech distortion into account in its optimization criterion, MWF is generalized as speech distortion weighted multichannel wiener filter (SDW-MWF), which provides a tradeoff between speech distortion and noise reduction \cite{souden2010optimal, spriet2004spatially, doclo2002gsvd}. In this work, a tradeoff parameter optimized method based on SDW-MWF is used.

Considering an array of $M$ microphones. Let $Y_m(k,l)$, $m=1,\dots,M$ denote the short-time Fourier transform (STFT) domain notation of $m$-th microphone signal at frequency index $k$ and frame index $l$, the received signals are given as
\begin{eqnarray}
Y_m(k,l) \!\!\!&=&\!\!\! S(k,l)G_m(k,l) + N_m(k,l)\nonumber  \\
		 \!\!\!&=&\!\!\! X_m(k,l) + N_m(k,l) \label{eq1}
\end{eqnarray}
where $S(k,l)$,$G_m(k,l)$,$X_m(k,l)$,$N_m(k,l)$ are respectively the STFT domain expression of the source signal $s(t)$, the transfer function from the source to the $m$-th microphone $g_m(t)$, the target signal $x_m(t)$ and noise signal $n_m(t)$ at microphone $m$.

To find an optimal estimate of the target signal, the designed SDW-MWF criterion is \cite{spriet2004spatially,doclo2007frequency}
\begin{eqnarray}
 \underset{\:\:\qquad\qquad\qquad\qquad\boldsymbol{w}}{\boldsymbol{w}_{SDW-MWF} =  \textrm{arg} \; \textrm{min}\;} E\{|\boldsymbol{w}^{H}\boldsymbol{y}-X_1|^2+\mu|\boldsymbol{w}^{H}\boldsymbol{n}|^2\}  \label{eq2}
\end{eqnarray}
where $X_1$ is the target signal at the first microphone, $\boldsymbol{y}(k,l)$ is the received signal vector defined as $\boldsymbol{y}(k,l)\\=[Y_1(k,l),\dots,Y_M(k,l)]^T$ and $\boldsymbol{w}^{H}(k,l)$, $\boldsymbol{x}(k,l)$, $\boldsymbol{n}(k,l)$, $\boldsymbol{g}(k,l)$ are defined similarly, among which $\boldsymbol{w}(k,l)$ represents the linear filter given by $\boldsymbol{w}(k,l)=[W_1(k,l),\dots,W_M(k,l)]^T$. Here operators $(.)^T$ and $(.)H$ represent the transposition and Hermitian transpose operation respectively.
Apparently, a larger value of $\mu$ emphasize more on noise reduction. Variables $k$ and $l$ are omitted here for simplicity. The solution to SDW-MWF can be obtained as
\begin{eqnarray}
\boldsymbol{w}_{SDW-MWF} = [\boldsymbol{\Phi}_{xx}+\mu\boldsymbol{\Phi}_{nn}]^{-1}\boldsymbol{\Phi}_{xx}\boldsymbol{u_1} \label{eq3}
\end{eqnarray}
where $\boldsymbol{u_1}=[1 \dots 0 \dots 0]^{T}$ is a $M$-dimensional vector corresponds to the first microphone (channel 1 of the 6-microphone array), $\boldsymbol{\Phi}_{xx}$ and $\boldsymbol{\Phi}_{nn}$ are the correlation matrices of clean speech signal and noise signal, respectively.

Using a fixed parameter $\mu$, the reduced residual noise level generally achieved at the expense of increased speech distortion. In our work, we compute the parameter according to desired noise reduction level.
\begin{eqnarray}
\mu = \textrm{min}(s,s/\rm{SNR}_i) \label{eq4}
\end{eqnarray}
where $\rm{SNR}_i$ denotes the imput signal-to-noise ratio (SNR) of the first microphone, $s$ is a noise reduction control factor defined as
$s=\phi_{n_1n_1}/\phi_0$, $\phi_{n_1n_1}$ represents the noise power at the first microphone, and $\phi_0$ represents desired residual noise level.
Apparently, when the background noise level is relatively high or the input SNR is relatively low, the optimized parameter will emphasize more on noise reduction, which is reasonable. In this work, the noise power and noise covariance matrix for each frequency bin are computed from the initial and final 10 frames of each utterance.

\section{BACK-END DESCRIPTION}

\subsection{Acousitic modeling with neural network}
\label{sec:modeling}
\begin{figure*}[htb]
  \centering
  \includegraphics[width=0.6\textwidth]{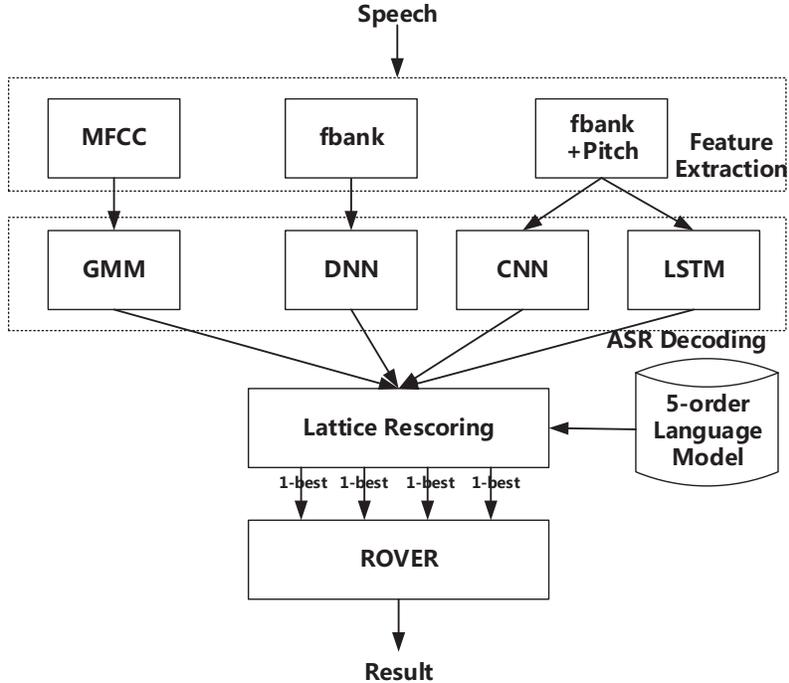}\\
  \caption{Back-end description}
  \label{fig:ASR}
\end{figure*}

Fig.\ref{fig:ASR} demonstrates the back-end description including the techniques we used of the proposed system.

The GMM baseline includes the standard triphone based acoustic models with various feature transformations including linear discriminant analysis (LDA), maximum likelihood linear transformation (MLLT), and feature space maximum likelihood linear regression (fMLLR) with speaker adaptive training (SAT).

The DNN baseline provides the state-of-the-art ASR performance. It is based on the Kaldi recipe for Track 2 of the 2nd CHiME Challenge \cite{weng2014recurrent}. The DNN is trained using the standard procedure (pre-training using restricted Boltzmann machine, cross entropy training, and sequence discriminative training). This baseline requires relatively massive computational resources (GPUs for the DNN training and many CPUs for lattice generation).

We start DNN training based on scripts of baseline system. We use 7 hidden layers and 2048 nodes for each hidden layer. The features for the DNN training are 40-dimensional filter-bank and its delta, delta-delta features. A context window of 11 frames (5+1+5) is used so that the dimension of the input layer for DNN is 40$*$3$*$11. Cepstral Mean and Variance Normalization (CMVN) is applied and proves to be useful. The DNN output layer size is the same as the GMM-HMM, which is 2024. The DNN is trained using the standard procedure like baseline system.

The CNN uses fbank+pitch features and contains two convolutional hidden layers and a max-pooling layer. The input feature vector (not including pitch) is divided into 40 bands. The corresponding dimension of the 11 consecutive feature frames are arranged in each band, together with their derivatives. So that the input dimension of the CNN is 43$*$3$*$11. The first set of convolutional filters are applied to 8 consecutive bands and generate 128 feature mappings. We then apply max-pooling across 3 bands to generate 11 bands. The second set of convolutional filters are applied to 4 consecutive bands and generate 256 feature mappings. Four fully-connected hidden layers of 1024 nodes are arranged after the convolutional layers. The total number of parameters for the CNN is 7.7M.

The LSTM network used in this paper is a two layer LSTM RNN, where each LSTM layer has 1024 memory cells and a dimensionality reducing recurrent projection layer of 200 linear units \cite{Sak2014Long, Sak2014Long2}.

In our experiments, we use an official trigram language model (LM) on the initial decoding pass and use a 5-order LM for lattice rescoring in a second pass. The official trigram LM has 5k vocabularies. The 5-order LM is trained using official training data only, but has vocabularies up to 12k.

\subsection{Combination of different systems}
\label{sec:rover}
To combine these multiple speech recognition outputs into a single one, we employ ROVER at the decision level \cite{Fiscus1997} in the final step. The fusion enables us to achieve a lower error rate than any of the individual systems alone. In this paper, NIST scoring toolkit (SCTK,version 1.3) is used as a rover tool to combine the different results. It takes N input files and does an N-way dynamic programming (DP) alignment on those files. The output is a voted output depending the maximum confidence score.

\section{EXPERIMENTS AND RESULTS}
\label{sec:experiments}

The experiments are all carried out following the instructions of CHiME challenge. In this section, we list the ASR improvement step by step according to each technique we used resulting in the final WER of the test set provided by CHiME challenge. Table.\ref{table:baseline} gives the GMM and DNN baselines 'CHiME' provided, Table.\ref{table:proposed} shows the ASR results by the proposed system and Table.\ref{table:scenes} shows the ASR results under each scenario including the bus (BUS), cafe (CAF), pedestrian area (PED), and street junction (STR) according to the best system after ROVER.
\begin{table*}[htb!]
\centering  
\begin{tabular}{|p{2.0cm}|p{2.0cm}|p{2.0cm}|p{1.0cm}|p{1.0cm}|p{1.0cm}|p{1.0cm}|}
\hline
\multicolumn{1}{|c|}{{\bf Model}}&\multicolumn{1}{|c|}{{\bf Test Data}}&\multicolumn{1}{|c|}{{\bf Training Data}}&\multicolumn{2}{|c|}{{\bf Dev. Set}}&\multicolumn{2}{|c|}{{\bf Test Set}}\\
\cline{4-7}
\multicolumn{1}{|c|}{{\bf}}&\multicolumn{1}{|c|}{{\bf}}&\multicolumn{1}{|c|}{{\bf}}&\multicolumn{1}{|c|}{{\bf Real}}&\multicolumn{1}{|c|}{{\bf Sim.}}&\multicolumn{1}{|c|}{{\bf Real}}&\multicolumn{1}{|c|}{{\bf Sim.}}\\
\hline
\multicolumn{1}{|c|}{{\bf}}&\multicolumn{1}{|c|}{{\bf noisy}}&\multicolumn{1}{|c|}{{\bf clean}}&55.65&50.25&79.84&63.30\\
\cline{3-7}
\multicolumn{1}{|c|}{{\bf GMM}}&\multicolumn{1}{|c|}{{\bf}}&\multicolumn{1}{|c|}{{\bf noisy}}&18.70&18.71&33.23&21.59\\
\cline{2-7}
\multicolumn{1}{|c|}{{\bf}}&\multicolumn{1}{|c|}{{\bf MVDR}}&\multicolumn{1}{|c|}{{\bf clean}}&41.88&21.72&78.12&25.63\\
\cline{3-7}
\multicolumn{1}{|c|}{{\bf}}&\multicolumn{1}{|c|}{{\bf}}&\multicolumn{1}{|c|}{{\bf MVDR}}&20.55&9.79&37.36&10.59\\
\hline
\multicolumn{1}{|c|}{{\bf DNN+sMBR}}&\multicolumn{1}{|c|}{{\bf noisy}}&\multicolumn{1}{|c|}{{\bf noisy}}&16.13&14.30&33.43&21.51\\
\hline
\multicolumn{1}{|c|}{{\bf DNN+sMBR}}&\multicolumn{1}{|c|}{{\bf MVDR}}&\multicolumn{1}{|c|}{{\bf MVDR}}&17.72&8.17&33.76&11.19\\
\hline
\end{tabular}
\caption{WER Baselines from the 3rd CHiME challenge.}
\label{table:baseline}
\end{table*}

\begin{table*}[htb!]
\centering  
\begin{tabular}{|p{2.0cm}|p{2.0cm}|p{2.0cm}|p{1.0cm}|p{1.0cm}|p{1.0cm}|p{1.0cm}|}
\hline
\multicolumn{1}{|c|}{{\bf Model}}&\multicolumn{1}{|c|}{{\bf Test Data}}&\multicolumn{1}{|c|}{{\bf Training Data}}&\multicolumn{2}{|c|}{{\bf Dev. Set}}&\multicolumn{2}{|c|}{{\bf Test Set}}\\
\cline{4-7}
\multicolumn{1}{|c|}{{\bf}}&\multicolumn{1}{|c|}{{\bf}}&\multicolumn{1}{|c|}{{\bf}}&\multicolumn{1}{|c|}{{\bf Real}}&\multicolumn{1}{|c|}{{\bf Sim.}}&\multicolumn{1}{|c|}{{\bf Real}}&\multicolumn{1}{|c|}{{\bf Sim.}}\\
\hline
\multicolumn{1}{|c|}{{\bf GMM}}&\multicolumn{1}{|c|}{{\bf SDW-MWF}}&\multicolumn{1}{|c|}{{\bf Clean}}&30.23&29.75&53.43&41.58\\
\cline{3-7}
\multicolumn{1}{|c|}{{\bf}}&\multicolumn{1}{|c|}{{\bf}}&\multicolumn{1}{|c|}{{\bf SDW-MWF}}&13.16&14.11&23.19&18.65\\
\hline
\multicolumn{1}{|c|}{{\bf GMM}}&\multicolumn{1}{|c|}{{\bf SDW-MWF}}&\multicolumn{1}{|c|}{{\bf Random SNR+SDW-MWF}}&13.01&13.95&22.07&17.57\\
\hline
\multicolumn{1}{|c|}{{\bf GMM+Rescore}}&\multicolumn{1}{|c|}{{\bf }}&\multicolumn{1}{|c|}{{\bf}}&11.61&12.37&20.35&15.7\\
\hline
\multicolumn{1}{|c|}{{\bf DNN+sMBR}}&\multicolumn{1}{|c|}{{\bf SDW-MWF}}&\multicolumn{1}{|c|}{{\bf Random SNR+SDW-MWF}}&9.95&10.03&18.4&12.98\\
\hline
\multicolumn{1}{|c|}{{\bf DNN+sMBR+Rescore}}&\multicolumn{1}{|c|}{{\bf}}&\multicolumn{1}{|c|}{{\bf}}&8.48&9.01&15.3&11.29\\
\hline
\multicolumn{1}{|c|}{{\bf CNN+sMBR}}&\multicolumn{1}{|c|}{{\bf SDW-MWF}}&\multicolumn{1}{|c|}{{\bf Random SNR+SDW-MWF}}&9.52&9.64&17.87&12.64\\
\hline
\multicolumn{1}{|c|}{{\bf CNN+sMBR+Rescore}}&\multicolumn{1}{|c|}{{\bf}}&\multicolumn{1}{|c|}{{\bf}}&8.51&8.77&16.37&11.55\\
\hline
\multicolumn{1}{|c|}{{\bf LSTM}}&\multicolumn{1}{|c|}{{\bf SDW-MWF}}&\multicolumn{1}{|c|}{{\bf Random SNR+SDW-MWF}}&10.81&11.18&18.96&14.1\\
\hline
\multicolumn{1}{|c|}{{\bf LSTM+Rescore}}&\multicolumn{1}{|c|}{{\bf}}&\multicolumn{1}{|c|}{{\bf}}&9.44&9.71&16.45&12.48\\
\hline
\multicolumn{1}{|c|}{{\bf ROVER}}&\multicolumn{1}{|c|}{{\bf SDW-MWF}}&\multicolumn{1}{|c|}{{\bf Random SNR+SDW-MWF}}&{\bf7.29}&{\bf7.68}&{\bf13.2}&{\bf9.71}\\
\hline
\end{tabular}
\caption{WERs of proposed system.}
\label{table:proposed}
\end{table*}

\subsection{ASR performance of front-end speech enhancement}
\label{sec:speech_enhancement}

As mentioned above, front-end speech enhancement brings benefits to the ASR performance. Table.\ref{table:proposed} demonstrates that WER of real test data decreases from 37.36\% to 23.19\% by changing the speech enhancement method from MVDR (supplied by CHiME organizers \cite{mestre2003diagonal}) to the proposed SDW-MWF under GMM acoustic model. If we randomize the SNR of training data from -6dB--6dB (denoted by {\bf Random SNR} in Table.\ref{table:proposed}) instead of the estimated SNR calculated from really recorded data for simulating training set, the WER decreases to 22.07\%.

Under DNN+sMBR acoustic model, the WER decreases from 33.76\% to 18.4\% on test data using SDW-MWF and random SNR schemes. It is worthy mentioning that all the training data is enhanced to compensate the mismatch between the training data and test data.

\subsection{Back-end ASR performance}

The results of DNN model on the development and evaluation set are also given in Table \ref{table:proposed}. we can see that DNN get 16.63\% relative WER reduction comparing with GMM system on the real data of the test set. Obviously, the improvement is not enough, then we tried to use several other NN topologies.

As it is shown in Table \ref{table:proposed}, the CNN acoustic models as it has shown superior performance over conventional DNN. The WER decreases from 18.4\% to 17.87\%. Table \ref{table:proposed} shows that LSTM gets further improvement. 14.09\% relative reduction was achieved comparing to GMM. After lattice rescoring, all of the systems get significantly improvement.

Finally the best ASR result was obtained by combining all the systems with lattice rescoring together. We achieve a final WER of 13.2\% on the real data of the test set, resulting in a 60.9\% relative reduction in WER compared to the result of 33.23\% from the best GMM-baseline. Table.\ref{table:scenes} shows the detail ASR results under different recording scenarios.

The best single system is the DNN+sMBR using lattice rescoring shown by Table.\ref{table:proposed}.

\begin{table}[htb!]
\centering  
\begin{tabular}{|p{2.0cm}|p{1.0cm}|p{1.0cm}|p{1.0cm}|p{1.0cm}|}
\hline
\multicolumn{1}{|c|}{{\bf Environment}}&\multicolumn{2}{|c|}{{\bf Dev. Set}}&\multicolumn{2}{|c|}{{\bf Test Set}}\\
\cline{2-5}
\multicolumn{1}{|c|}{{\bf}}&\multicolumn{1}{|c|}{{\bf Real}}&\multicolumn{1}{|c|}{{\bf Sim.}}&\multicolumn{1}{|c|}{{\bf Real}}&\multicolumn{1}{|c|}{{\bf Sim.}}\\
\hline
\multicolumn{1}{|c|}{{\bf BUS}}&8.88&6.77&17.74&7.4\\
\hline
\multicolumn{1}{|c|}{{\bf CAF}}&7.08&9.94&11.75&10.95\\
\hline
\multicolumn{1}{|c|}{{\bf PED}}&5.78&6.14&13.34&9.19\\
\hline
\multicolumn{1}{|c|}{{\bf STR}}&7.4&7.89&9.96&11.32\\
\hline
\end{tabular}
\caption{WERs of the best system under different environments.}
\label{table:scenes}
\end{table}

%


\section{CONCLUSION}
\label{sec:conclusion}
A state-of-the-art ASR system is presented in this paper facing with the task of reducing the effects of noise under different real applicable scenarios using a 6-microphone array. Two aspects are stated separately. Front-end speech enhancement using SDW-MWF achieves considerable performance improvement. Back-end techniques including GMM, DNN, CNN and LSTM are investigated. The combination of the four systems with lattice rescoring has the best ASR performance on the develop and test set. we achieve a relative 60.9\% WER reduction on the real data of the test data compared to the best baseline system.

\bibliographystyle{IEEEbib}
\bibliography{strings,refs}

\end{document}